\documentclass[journal=jacsat,manuscript=article]{achemso}

\usepackage{chemformula} 
\usepackage[T1]{fontenc} 
\usepackage{algorithm2e}
\usepackage{subfig}

\author{Boqin Zhang}
\affiliation[CEEgatech]
{College of Engineering, Georgia Institute of Technology, Atlanta, GA, USA}
\author{Shikhar Shah}
\affiliation[CSEgatech]
{College of Computing, Georgia Institute of Technology, Atlanta, GA, USA}
\author{John E. Pask}
\affiliation{Physics Division, Lawrence Livermore National Laboratory, Livermore, CA, USA}
\author{Edmond Chow}
\affiliation[CSEgatech]
{College of Computing, Georgia Institute of Technology, Atlanta, GA, USA}
\author{Phanish Suryanarayana}
\affiliation[CEEgatech]
{College of Engineering, Georgia Institute of Technology, Atlanta, GA, USA}
\alsoaffiliation[CSEgatech]
{College of Computing, Georgia Institute of Technology, Atlanta, GA, USA}
\email{phanish.suryanarayana@ce.gatech.edu}

\title{Random Phase Approximation Correlation Energy using Real-Space Density Functional Perturbation Theory}

\begin{document}

%
%

\begin{abstract}
We present a real-space method for computing the random phase approximation (RPA) correlation energy within Kohn–Sham density functional theory, leveraging the low-rank nature of the frequency-dependent density response operator. In particular, we employ a cubic scaling formalism based on density functional perturbation theory that circumvents the calculation of the response function matrix, instead relying on the ability to compute its product with a vector through the solution of the associated Sternheimer linear systems. We develop a large-scale parallel implementation of this formalism using the subspace iteration method in conjunction with the spectral quadrature method, while employing the Kronecker product-based method for the application of the Coulomb operator and the conjugate orthogonal conjugate gradient method for the solution of the  linear systems. We demonstrate convergence with respect to key parameters and verify the method's accuracy by comparing with planewave results. We show that the framework achieves good strong scaling to many thousands of processors, reducing the time to solution for a lithium hydride system with 128 electrons to around 150 seconds  on 4608 processors. 
\end{abstract}

\section{Introduction}

Over the past few decades, quantum mechanical calculations have become indispensable in materials and chemical sciences research, providing both fundamental insights and predictive power. Among the various first principles approaches, Kohn–Sham density functional theory (DFT) \cite{Kohn1965, hohenberg1964inhomogeneous} has emerged as one of the most widely used, largely due to its versatility, relative simplicity, and favorable accuracy-to-cost ratio. Nevertheless, solving the Kohn–Sham equations remains computationally demanding, imposing significant limitations on the size and complexity of systems as well as the time scales that can be explored. These challenges become especially acute with the choice of more advanced exchange-correlation functionals. 

The exchange-correlation functional in Kohn–Sham DFT is used to model both electron exchange, a quantum mechanical effect enforcing the Pauli exclusion principle, and electron correlation, which captures the dynamic interactions between electrons. In particular, exchange-correlation functionals can be classified by their accuracy and complexity within the conceptual framework of Jacob’s Ladder \cite{perdew2001jacob}, where each higher rung represents a more advanced and generally more accurate representation of  exchange and correlation. The fifth and highest rung includes the random phase approximation (RPA) correlation energy, which incorporates many-body effects and can be derived from the adiabatic connection fluctuation dissipation (ACFD) theorem \cite{langreth1977exchange},  ideally used alongside exact exchange. It can capture van der Waals interactions, eliminate self-interaction errors, and is applicable to both small-gap and metallic systems, enabling benchmark results for condensed matter systems \cite{ren2012random, eshuis2012electron}. In particular, it has been found to offer better predictive capabilities than lower rung  functionals for a range of properties, including surface energy, adsorption energy, binding energy, cohesive energy, and lattice constants \cite{ren2009exploring, harl2010assessing, del2015probing, cui2016first, hermann2017first, schmidt2018benchmark,sheldon2023adsorption, oudot2024reaction}. 

The RPA correlation energy is expressed in terms of the Coulomb operator and the non-interacting Kohn–Sham density response function at imaginary frequency, which depends on both the occupied and unoccupied orbitals unlike lower-rung functionals that require only the occupied orbitals. Standard RPA correlation energy calculations \cite{gonze2016recent, Enkovaara2010, VASP} exhibit quartic scaling with the number of grid points, and consequently with system size, while being associated with a very large computational prefactor. As a result, RPA calculations can be orders of magnitude more expensive than commonly used local/semilocal exchange-correlation functionals. This has motivated the development of approaches with reduced prefactor and/or scaling \cite{nguyen2009efficient, rocca2014random, del2013electron, kaltak2014low, kaltak2014cubic, nguyen2014ab, wilson2008efficient, wilson2009iterative, qin2023interpolative, weinberg2024static}, as well as  efficient/scalable parallel implementations \cite{del2015enabling, qin2023interpolative, weinberg2024static}.  However, these frameworks are based on the planewave method \cite{martin2020electronic}, which confines calculations to periodic boundary conditions due to the underlying Fourier basis, necessitating artificial periodicity with large vacuum regions for finite systems such as molecules and clusters, as well as for semi-infinite systems like surfaces and nanotubes. Additionally, a neutralizing background density is required to prevent Coulomb divergences when treating charged systems and the method’s reliance on fast Fourier transforms (FFTs) can hamper scalability on large-scale parallel computing platforms. 

In view of the limitations of the planewave method, various approaches using systematically improvable, localized representations have been developed over the past two decades \cite{chelikowsky1994finite, genovese2008daubechies, white1989finite, tsuchida1995electronic, xu2018discrete, suryanarayana2011mesh, suryanarayana2010non, ONETEP, CONQUEST, MOTAMARRI2020106853, castro2006octopus, briggs1996real, fattebert1999finite, ghosh2017sparc, pask2005femeth, lin2012adaptive}. Among these, finite-difference methods \cite{saad2010esmeth} stand out as perhaps the most mature and widely used to date. By discretizing all relevant quantities on a real-space grid, these methods maximize computational locality while accommodating Dirichlet as well as periodic/Bloch-periodic boundary conditions, and combinations thereof. This capability allows for the efficient and accurate treatment of finite, semi-infinite, and  bulk 3D materials. Additionally, convergence is governed by a single parameter, i.e., grid spacing, and the method’s inherent simplicity and locality, along with its avoidance of communication-intensive transforms like FFTs, enable efficient scaling on large-scale parallel computing platforms. In particular, these methods can significantly out-perform their planewave counterparts using local, semilocal, and hybrid exchange-correlation functionals, with increasing advantages as the number of processors is increased \cite{xu2021sparc, ZHANG2024100649, jing2024efficient}. Furthermore, they are capable of exploiting the decay of electronic interactions with distance, which has enabled the study of very large systems containing a million atoms for local/semilocal exchange-correlation \cite{fattebert2016modeling, gavini2022roadmap}. However, the RPA correlation energy has not been implemented within the real-space method heretofore,  to our knowledge, which provides the motivation for the present work. 

In this work, we develop a real-space framework for calculating the RPA correlation energy within Kohn–Sham DFT, leveraging the low-rank nature of the frequency-dependent density response operator to avoid the explicit construction of the full response matrix. In particular, we employ  a density functional perturbation theory (DFPT) \cite{baroni2001phonons, gonze1989density}-based formalism for evaluation of the matrix–vector products via the Sternheimer linear systems, reducing the overall scaling to cubic. We develop a highly scalable parallel implementation based on the subspace iteration \cite{saad2011numerical} and spectral quadrature (SQ) \cite{suryanarayana2013spectral, suryanarayana2013coarse} methods, while employing a Kronecker product–based scheme \cite{jing2024efficient, jing2025gpu} for application of the Coulomb operator and the conjugate orthogonal conjugate gradient (COCG) method \cite{van1990petrov, Shah2024Many} for solution of the linear systems. We demonstrate convergence with respect to key parameters, verify the method's accuracy by comparison with planewave results, and show that the framework achieves excellent strong scaling to many thousands of processors.

The remainder of this paper is organized as follows. First, we discuss the DFPT-based approach for computing the RPA correlation energy. Next, we describe its implementation within the open-source SPARC electronic structure code \cite{xu2021sparc, ZHANG2024100649}  and evaluate its accuracy and performance. Finally, we provide concluding remarks and outline potential directions for future work.


\section{Formulation \label{Sec:Formulation}} 
The RPA correlation energy can be written as \cite{niquet2003exchange}:
\begin{equation}
	\label{eqn:rpa-energy}
	E_c = \frac{1}{2\pi} \int_0^\infty \operatorname{Tr}[\log(\mathcal{I} - \chi_0(i\omega)\nu) + \chi_0(i\omega)\nu] \, \mathrm{d\omega} \,,
\end{equation}
where $ \rm Tr[\cdot]$ represents the trace operator, $\log$ denote the natural logarithm,  $\mathcal{I}$ is the identity operator, $\chi_0(i\omega)$ is the non-interacting Kohn-Sham density response function at imaginary frequency $i \omega$, and $\nu$ is the Coulomb operator. Considering isolated systems or extended systems with $\Gamma$-point Brillouin zone integration, the response function for closed shell systems when neglecting spin can be written as \cite{adler1962quantum, wiser1963dielectric}: 
\begin{equation}
	\label{eqn:chi0def}
	\chi_0(r, r'; i\omega) = 2 \sum_{j}\sum_{k} \frac{(f_j - f_k)  \psi_j(r) \psi_k(r) \psi_k(r') \psi_j(r')}{\varepsilon_j - \varepsilon_k - i\omega} \,,
\end{equation}
where $\psi$ and $\varepsilon$ are the eigenfunctions (orbitals) and eigenvalues of the Hamiltonian $\mathcal{H}$:
\begin{align}
\mathcal{H} \psi_n = \varepsilon_{n} \psi_n \,,
\end{align}
the sums over the indices \( j \) and \( k \)  run over both the occupied and unoccupied orbitals, and $f \in \{0,1\}$ are the occupations such that $2 \sum_{n} f_n = N_e$, $N_e$ being the total  number of electrons. In pseudopotential real-space density functional theory, the Hamiltonian takes the form \cite{ghosh2017sparc, ghosh2017sparcb}:
\begin{align}
\mathcal{H} :=  -\frac{1}{2} \nabla^2 + V_{xc} + \phi  + V_{nl} \,,
\end{align}
where $\nabla^2$ denotes the Laplacian, $V_{xc}$ is the exchange-correlation potential, $V_{nl}$  is the nonlocal pseudopotential operator, and $\phi$ is the electrostatic potential, which is the solution to the Poisson equation:
\begin{align}
- \frac{1}{4 \pi} \nabla^2 \phi = \rho + b \,,
\end{align}
with $\rho$ and $b$ being the electron and pseudocharge densities, respectively. Note that since $\chi_0$ is a negative definite operator, the RPA correlation energy is well-defined in all instances. 

The RPA correlation energy in Eq.~\ref{eqn:rpa-energy} can be evaluated within the real-space method as:
\begin{align}
E_c \approx \frac{1}{2\pi}  \int_{0}^{\infty}  \sum_{n=1}^{N_{d}} \left( \log(1 - \lambda_n(i \omega)) + \lambda_n(i \omega) \right)  \, \mathrm{d} \omega \,,
\end{align}
where $N_d$ is the number of grid points, and $\lambda_n$ are the eigenvalues of the $\chi_0 \nu$-matrix, or equivalently those of the $\widetilde{\chi}_0 = \nu^{1/2} \chi_0 \nu^{1/2}$-matrix, which shares the same eigenvalues as $\chi_0 \nu$, but is generally more efficient and convenient for use in computations. At any given frequency, the contribution to the correlation energy can be determined by first calculating the $\chi_0$-matrix,  and then the $\widetilde{\chi}_0$-matrix, followed by a  computation of its eigenvalues. The calculation of the $\chi_0$-matrix using Eq.~\ref{eqn:chi0def} scales as $\mathcal{O}(N_d^4)$ with respect to the number of grid points, $N_d$, while its computer storage scales quadratically, $\mathcal{O}(N_d^2)$, leading to the overall RPA correlation energy calculation having the same scaling behavior. Given that the number of grid points typically ranges from 400 to 30,000 per atom \cite{xu2018discrete},  the calculation of the RPA correlation energy is prohibitively expensive, restricting such simulations to particularly small  systems. Furthermore, the calculation of the $\chi_0$-matrix requires the unoccupied orbitals, which are not readily available from standard Kohn-Sham calculations.

\begin{figure}[htbp!]
\centering
\includegraphics[keepaspectratio=true,width=0.99\textwidth]{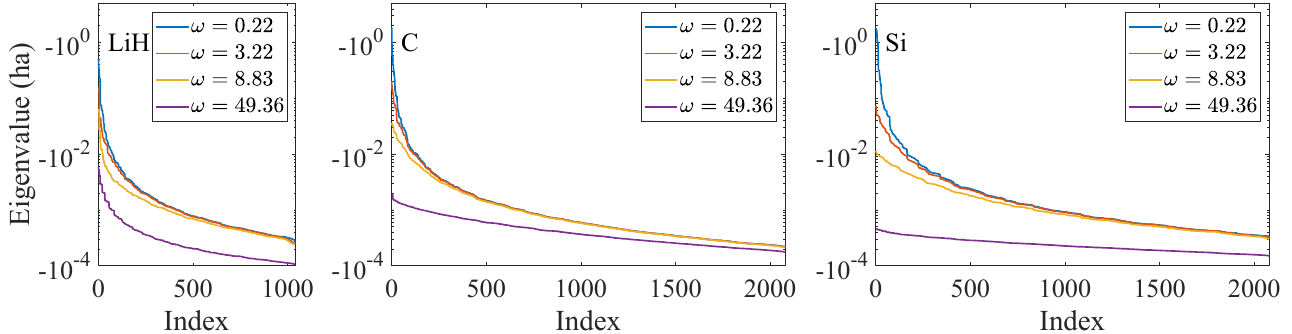}
\caption{Decay in the eigenvalues of the $\widetilde{\chi}_0$-matrix for 2-atom cells of lithium hydride (left), carbon (middle), and silicon (right) with atoms randomly perturbed, $\Gamma$-point Brillouin zone integration, and Perdew-Burke-Ernzerhof (PBE) \cite{perdew1996generalized} exchange-correlation functional.}
\label{Fig:Eigenvalues}
\end{figure}

In view of the above, it is preferable to use an iterative eigensolver that avoids the need to store the $\chi_0$ or $\widetilde{\chi}_0$ matrices, requiring only the evaluation of matrix-vector products.  Moreover, when computing such products, it is desirable to require only the occupied Kohn-Sham orbitals and eigenvalues. This can be accomplished within the framework of DFPT \cite{baroni2001phonons, gonze1989density}. In particular, it follows from the definition of the density response function that for any perturbation in the potential $\Delta V$:
\begin{align}
\chi_0 (\Delta V) = \Delta \rho \,,
\end{align}
where $\Delta \rho$ is the corresponding perturbation in the electron density, which can itself be written as:
\begin{align}
\Delta \rho = 4 \sum_{n} \Re [\psi_n (\Delta \psi_n)] \,,
\end{align}
with $\Delta \psi_n$ being the perturbation in the orbitals and $\Re$ representing the real part of the complex-valued quantity. The perturbation in the orbitals can be written as the solution to the Sternheimer equation \cite{sternheimer1954electronic, nguyen2014ab, nguyen2009efficient, wilson2008efficient, wilson2009iterative}:
\begin{align}
\left( \mathcal{H} - \varepsilon_n \mathcal{I} - i \omega \mathcal{I} \right) (\Delta \psi_n) = - \psi_n (\Delta V) \,, \quad n = 1, 2, \ldots N_s \,,
\end{align}
where $N_s$ is the number of occupied states. Given that the Hamiltonian is symmetric, the operator/matrix for this linear system is complex symmetric. To enable the use of efficient subspace-based iterative eigensolvers without being constrained by storage limitations, we assume a low-rank decomposition of the $\widetilde{\chi}_0$-matrix \cite{nguyen2014ab, nguyen2009efficient, wilson2008efficient, wilson2009iterative}, whereby the RPA correlation energy is approximated as:
\begin{align}
\label{Eq:RPA:LowRank}
E_c \approx \frac{1}{2\pi}  \int_{0}^{\infty}  \sum_{n=1}^{N_{r}} \left( \log(1 - \lambda_n(i \omega)) + \lambda_n(i \omega) \right)  \, \mathrm{d} \omega \,,
\end{align}
$N_{r}$ being the rank of the decomposition. Indeed, the eigenvalues of the $\widetilde{\chi}_0$-matrix rapidly decay to zero, as shown in Fig.~\ref{Fig:Eigenvalues}. Furthermore, under the transformation $g(x) = \log(1-x) + x$, eigenvalues near zero get closer to zero, whereby their contribution to the RPA correlation energy is diminished. Also, the  lower frequencies contribute more significantly to the correlation energy.

In summary, while calculating the eigenvalues of interest, i.e., the lowest $N_r$ eigenvalues of $\widetilde{\chi}_0$, the product of the $\chi_0$-matrix with any $\Delta V$ vector can be calculated by solving the Sternheimer equations for the different $\Delta \psi$ vectors, which are then used to calculate the $\Delta \rho$ vector. In this way, the calculation of the unoccupied states is avoided, and the formalism has a reduced scaling of $\mathcal{O}(N_r N_s N_d) \sim \mathcal{O}(N_d^3)$, while being accompanied by a relatively small prefactor. It is worth noting that by choosing the standard basis vectors for $\Delta V$, the DFPT-based formalism can also be used to calculate the $\chi_0$-matrix with $\mathcal{O}(N_s N_d^2) \sim \mathcal{O}(N_d^3)$ scaling, rendering the overall scaling to also be $\mathcal{O}(N_d^3)$, although this approach still requires storing the complete $\chi_0$-matrix and has a substantially larger prefactor.


\section{Implementation}

We now discuss the implementation of the DFPT-based formalism for the calculation of the  RPA correlation energy  within the SPARC \cite{xu2021sparc, ZHANG2024100649} electronic structure code. SPARC is based on the real-space finite-difference method, wherein high-order centered finite differences are used to approximate derivative operators, and the trapezoidal rule is applied for spatial integrations. It utilizes norm-conserving pseudopotentials and a local formulation for the  electrostatics.  The sparse Hamiltonian matrix is not explicitly computed or stored; instead, it is applied as an operator using a matrix-free scheme. In this work, we focus on the non self-consistent RPA correlation energy, which is calculated at the electronic ground state of a lower rung exchange-correlation functional. The ground state orbitals and eigenvalues, computed using SPARC, are stored in files and subsequently read during the  calculation of the correlation energy.

The pseudocode for the RPA correlation energy calculation is presented in Algorithm~\ref{alg:framework}, where the scaling of the key computational operations has also been listed. The integral over the frequency is approximated using Gauss-Legendre quadrature. At each frequency in the quadrature rule, the contribution to the RPA correlation energy is determined by using the subspace iteration method \cite{saad2011numerical},  a generalization of the power method, in conjunction with the Gauss Spectral Quadrature (SQ) method \cite{suryanarayana2013spectral, suryanarayana2013coarse}. The loop over the frequency proceeds from highest to lowest, since the solution of the Sternheimer equations becomes more challenging at the lower frequencies, given that the coefficient matrix becomes more ill-conditioned as $\omega \rightarrow 0$. In particular, starting with higher frequencies allows the eigenvector subspace of $\widetilde{\chi}_0$ computed at a given frequency to serve as a good initial guess for the subsequent lower frequency, reducing the number of iterations in the subspace iteration method and therefore the overall computational cost. At each frequency, the convergence of the associated RPA correlation energy  is used as the stopping criterion for the subspace iteration method. 

In the subspace iteration method, the $\widetilde{\chi}_0$-matrix is multiplied by trial vectors that, upon convergence, span the eigenspace corresponding to the lowest $N_r$ eigenvalues. For numerical stability, the trial vectors are orthogonalized in each iteration using the Cholesky factor of the overlap matrix, ensuring that the vectors do not become linearly dependent, an issue that arises due to the tendency of a power-like method to converge to the dominant eigenvalue. Note that in recent work \cite{Shah2024Many}, the subspace iteration method was used with second-degree polynomial filtering, requiring three $\widetilde{\chi}_0$-matrix products per iteration, whereas the current methodology requires only a single $\widetilde{\chi}_0$-matrix product per iteration, making the current implementation significantly faster. The RPA correlation energy at the given frequency is calculated using the Gauss SQ method applied to the $\widetilde{\chi}_0$-matrix projected into the subspace of the trial vectors, while not employing any truncation of the off-diagonal components. In particular, starting with each of the standard basis vectors in the subspace spanned by the trial vectors, the Lanczos iteration is used to generate the corresponding  tridiagonal matrix. The eigenvalues and square of the first components of the eigenvectors of this matrix provide the nodes and weights for the quadrature rule.  The choice of Gauss SQ rather than subspace diagonalization is motivated by its highly scalable nature \cite{suryanarayana2018sqdft, xu2022real}. In addition, the prefactor associatd with the method is expected to be relatively small due to the nature of the function being integrated, namely $g(x) = \log(1-x) + x$, for which quadrature orders as low as 3 are sufficiently accurate in practice.

The multiplication of the $\widetilde{\chi}_0$-matrix with any trial vector proceeds as follows. First, the product of the $\nu^{1/2}$-matrix with the vector is calculated using the real-space Kronecker product-based formalism \cite{jing2024efficient, jing2025gpu}, which has comparable efficiency to the fast Fourier transform (FFT) method, without the restriction of periodic boundary conditions. Next, the product of the $\chi_0$-matrix with the resulting vector is evaluated using the DFPT formalism described in the previous section. In particular, the Sternheimer equation is solved using the conjugate orthogonal conjugate gradient (COCG) method \cite{van1990petrov, Shah2024Many}, which represents a generalization of the conjugate gradient (CG) method to complex-symmetric linear systems.    The initial guess for the linear solver is constructed using Galerkin projection, where the components corresponding to the occupied Kohn-Sham states are removed from the initial residual \cite{Shah2024Many}. Note that though the solution of the linear system corresponding to the previous larger frequency likely serves as a good initial guess for the subsequent lower frequency, this results in a significant increase in the computer memory required, hence the strategy is not adopted here. Also note that due to load imbalance issues in parallel computations, i.e., the  Sternheimer linear system solution is more challenging for eigenvectors corresponding to lower eigenvalues of the $\widetilde{\chi}_0$-matrix, the trial vectors were cyclically reordered by multiplying with a permutation matrix in recent work \cite{Shah2024Many}. However, we found that this strategy does not provide any gains here, as the trial vectors are not the eigenvectors themselves, but merely span the same subspace, a consequence of using the SQ method rather than an eigensolver for the subspace eigenproblem. 

The implementation applies different parallelization strategies to each step, depending on the operations to be carried out in that step. In particular, when evaluating the $\widetilde{\chi}_0$-matrix product with the trial vectors, the multiplication of the $\nu^{1/2}$ matrix with the vectors uses a one-level parallelization over the different vectors, while the $\chi_0$-matrix product with the vectors employs a  two-level parallelization, first over the trial vectors and then over the Sternheimer linear systems. After evaluating the matrix-vector products, both the resulting matrix and the trial vector matrix are redistributed into block-cyclic form, with their multiplication performed using ScaLAPACK \cite{slug} routines on the subset of processors over which the trial vectors are parallelized. The resulting subspace matrix is then redistributed across a two-level group of processors, with each subgroup storing the matrix partitioned row-wise among the processors within that subgroup. The Gauss SQ implementation uses two levels of parallelization: first, across the different standard basis vectors within the processor group, and second, over the matrix-vector multiplications involved in the Lanczos iteration within the processor subgroup. After the SQ process, the orthogonalization of the $\widetilde{\chi}_0$-matrix multiplied vectors, i.e., calculation of the overlap matrix, Cholesky factorization, and subspace rotation, are all performed using ScaLAPACK  routines. The layout of the trial vectors is then restored to that used for the product of the $\widetilde{\chi}_0$-matrix with the trial vectors.

The key computational step in the above methodology is the solution of the  Sternheimer linear systems, the cost of which scales as $\mathcal{O}(N_d)$ each. Since the product of the $\chi_0$-matrix with each vector involves $N_s$ such linear systems, and assuming the number of iterations in the subspace iteration method is independent of the system size, the total number of matrix-vector products is $\mathcal{O}(N_r)$, leading to an overall scaling of $\mathcal{O}(N_r N_s N_d) \sim \mathcal{O}(N_d^3)$. Indeed, the generation of the initial guess for each linear system scales as $\mathcal{O}(N_s N_d)$, which leads to an overall scaling of $\mathcal{O}(N_r N_s^2 N_d) \sim \mathcal{O}(N_d^4)$.  However, for small to moderate system sizes, this cost constitutes only a small fraction of the total cost, making the implementation effectively cubic scaling in practice. To circumvent the need for the quartic scaling initial guess calculation,  we have also implemented Laplacian-based preconditioning based on the Kronecker product formalism \cite{jing2024efficient} and the block variant of the COCG linear solver \cite{Shah2024Many}. Indeed, preliminary results indicate that these modifications yield performance comparable to the current framework for small- to moderate-sized systems, while preserving the overall cubic scaling, as the generation of the initial guess can be omitted due to its marginal effect in this setting \cite{shah2024Thesis}.

\RestyleAlgo{ruled}
\SetKwComment{Comment}{$\triangleright\;$}{}
\begin{algorithm}[htbp!]
\caption{Pseudocode for the calculation of the RPA correlation energy, along with the scaling of the key computational operations.}\label{alg:framework}
$E_c$: RPA correlation energy, $\varepsilon_{c}$: threshold parameter for the RPA correlation energy \\ 
$H$: Hamiltonian, a matrix of dimension $N_d \times N_{d}$ applied as an operator \\
$P$:  Kohn-Sham orbitals, a matrix of dimension $N_d \times N_{s}$  \\
$V$:  trial vectors that span desired subspace of $\widetilde{\chi}_0$, a matrix of dimension $N_d \times N_{r}$ \\
$N$: $\nu^{1/2}$, a matrix of dimension $N_d \times N_{d}$ applied using Kronecker product formalism \\
$L$: lower triangular matrix, a matrix of dimension $N_r \times N_r$ \\ 
$I$: identity matrix of dimension $N_d \times N_{d}$, $O$: zero vector of dimension $N_d \times 1$ \\
$N_\omega$: frequency quadrature order,  $w$: frequency quadrature weights \\
$N_r$: rank of the decomposition for $\widetilde{\chi}_0$ \\
$N_s$: number of occupied Kohn-Sham orbitals \\ 
$N_d$: number of real space grid points \\ 
$N_o$: SQ order, $J$: tridiagonal matrix of dimension $N_{o} \times N_{o}$, $e_r$: standard basis vector \\ 
- - - - - - - - - - - - - - - - - - - - - - - - - - -  - - - - - - - - - - - - - - - - - - - - - - - - - - - - - - - -  \\
 $E_c = 0$ ; \\
\For (\tcp*[f]{\footnotesize Loop over frequency}){$m = N_\omega$  \KwTo $1$}  
{
$E_{c,m} = 0$ ;

\While (\tcp*[f]{\footnotesize Subspace iteration}){$|E_{c,m} - E_{c,m}^{old}| > \varepsilon_{c}/N_\omega$} 
{
$E_{c,m}^{old} = E_{c,m}$ ; \\
\For(\tcp*[f]{\footnotesize Loop over the trial vectors}){$j=1$ \KwTo $N_{r}$} 
{
$V_j = V(:,j)$ ; \\
$V_j = N V_j$ ;  \hspace{20mm} \tcp*[f]{\footnotesize $\nu^{1/2}$-matrix product with a vector} \Comment*[r]{$\mathcal{O}(N_d^{4/3})$}  
$R_j  = O$ ; \\
\For (\tcp*[f]{\footnotesize $\chi_0$-matrix product with a vector}){$n=1$ \KwTo $N_{s}$} 
{
$P_n = P(:,n)$ ; \\
$(H - \varepsilon_n I - i \omega_m) P_{j,n} = - P_n \odot V_j$ \,; \tcp*[f]{\footnotesize Sternheimer equation} \Comment*[r]{$\mathcal{O}(N_d)$} 
$R_j = R_j + 4 P_n \odot P_{j,n}$ \Comment*[r]{$\mathcal{O}(N_d)$} 
} 
$R_j = N R_j$ \,; \hspace{20mm} \tcp*[f]{\footnotesize $\nu^{1/2}$-matrix product with a  vector} \Comment*[r]{$\mathcal{O}(N_d^{4/3})$}  
$R(:,j) = R_j$ ; \\ 
} 
$C = V^{T} R$ \,; \hspace{65mm} \tcp*[f]{\footnotesize Projection} \Comment*[r]{$\mathcal{O}(N_{r}^2 N_d)$}
\For (\tcp*[f]{\footnotesize SQ method}){$r=1$ \KwTo $N_{r}$}
{
$L(:,1)= e_r$ ; \\
$C \approx L J L^T$    \Comment*[r]{$\mathcal{O}(N_{r}^2)$}
$J Q = Q \Lambda $ ; \\
$E_{c,m} = E_{c,m} + \sum_{t=1}^{N_{o}} Q(1,t)^2 (\ln (I-\mathit{\Lambda}(t,t)) + \mathit{\Lambda}(t,t))$ ;
} 
$M = R^{T} R$ \,; \hspace{30mm} \tcp*[f]{\footnotesize Overlap matrix computation} \Comment*[r]{$\mathcal{O}(N_{r}^2 N_d)$}
$L L^{T}=M$\,; \hspace{19mm} \tcp*[f]{\footnotesize Cholesky factorization} \Comment*[r]{$\mathcal{O}(N_{r}^3)$}
$V = R L^{-T}$ \,; \hspace{13mm} \tcp*[f]{\footnotesize Orthonormalization} \Comment*[r]{$\mathcal{O}(N_{r}^2 N_d)$}
} 
$E_c = E_c + E_{c,m}$ ;
} 
\end{algorithm}

\newpage
\section{Results and Discussion}
We now study the accuracy and performance of the DFPT-based framework  developed in this work for the calculation of the RPA correlation energy. We consider cells of silicon (Si), carbon (C), and lithium hydride (LiH), with the atoms randomly perturbed, using the $\Gamma$-point for Brillouin zone integration.  In all simulations, including electronic ground state calculations, we use the Perdew-Burke-Ernzerhof (PBE) exchange-correlation functional \cite{perdew1996generalized} and ONCV pseudopotentials \cite{hamann2013optimized} with nonlinear core corrections from the SPMS table \cite{shojaei2023soft},  which have 4, 4, 3, and 1 electrons in valence for Si, C, Li, and H, respectively. In all instances, we set the order of the Gauss-Legendre quadrature  for the integral over the frequency as $N_\omega=8$. 


\subsection{Convergence with respect to parameters}
We first study the convergence of the RPA correlation energy with respect to the key parameters: normalized rank  of the decomposition $N_r/N_e$, where $N_e$ is the total number of electrons; tolerance $\varepsilon_s$ for solving the Sternheimer equation, prescribed on the relative residual; real-space grid spacing $h$; and SQ order $N_o$.  We consider 8-atom cubic cells with dimensions of 10.3, 6.7, and 7.6 bohr for the Si, C, and LiH systems, respectively, containing 32, 32, and 16 electrons. Unless specified otherwise, we employ $N_o=7$, $\varepsilon_s=10^{-3}$, $N_r/N_e = 260$, subspace iteration RPA correlation energy threshold of $\varepsilon_c=10^{-6}$ ha/atom, and $h=0.20$, $0.15$, and $0.15$ bohr for the Si, C, and LiH systems, respectively, resulting in $N_d= \text{140,608}$, $\text{91,125}$, and $\text{132,651}$ grid points. 

\begin{figure}
\centering
\subfloat[Nomalized decomposition rank]{\includegraphics[keepaspectratio=true,width=0.45\textwidth]{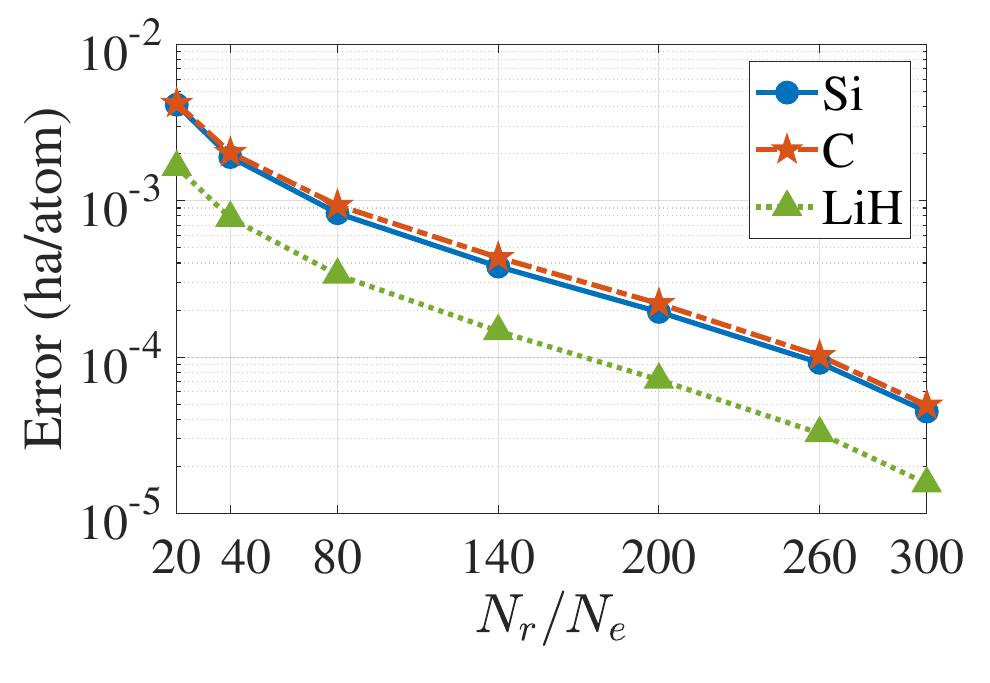} \label{fig:conv:rank}} 
\subfloat[Sternheimer tolerance]{\includegraphics[keepaspectratio=true,width=0.45\textwidth]{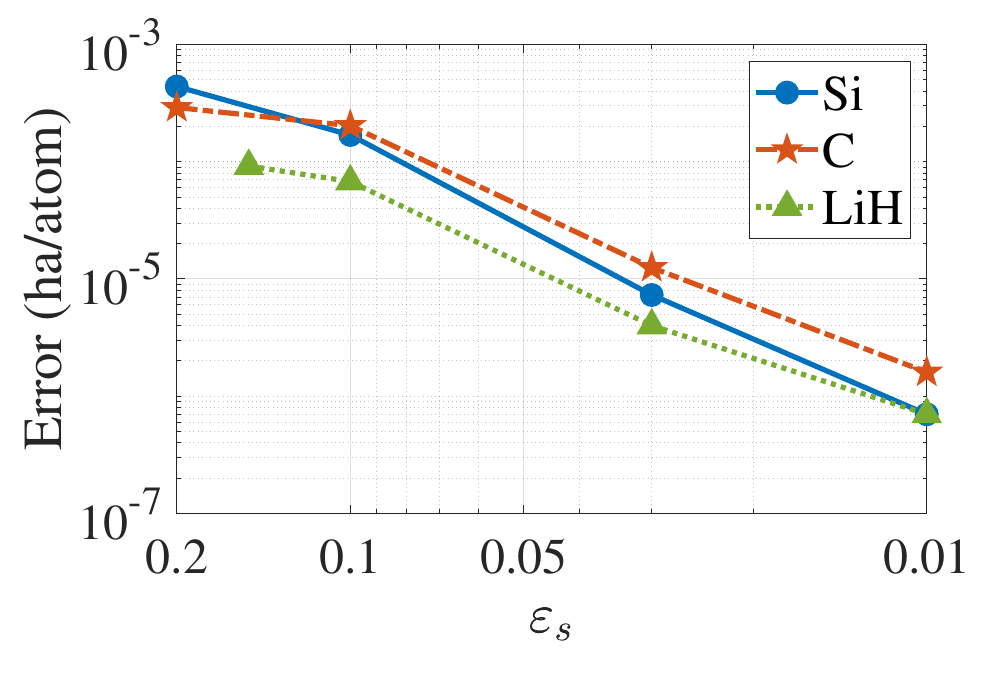}  \label{fig:conv:SternTol}} \\
\subfloat[Grid spacing]{\includegraphics[keepaspectratio=true,width=0.45\textwidth]{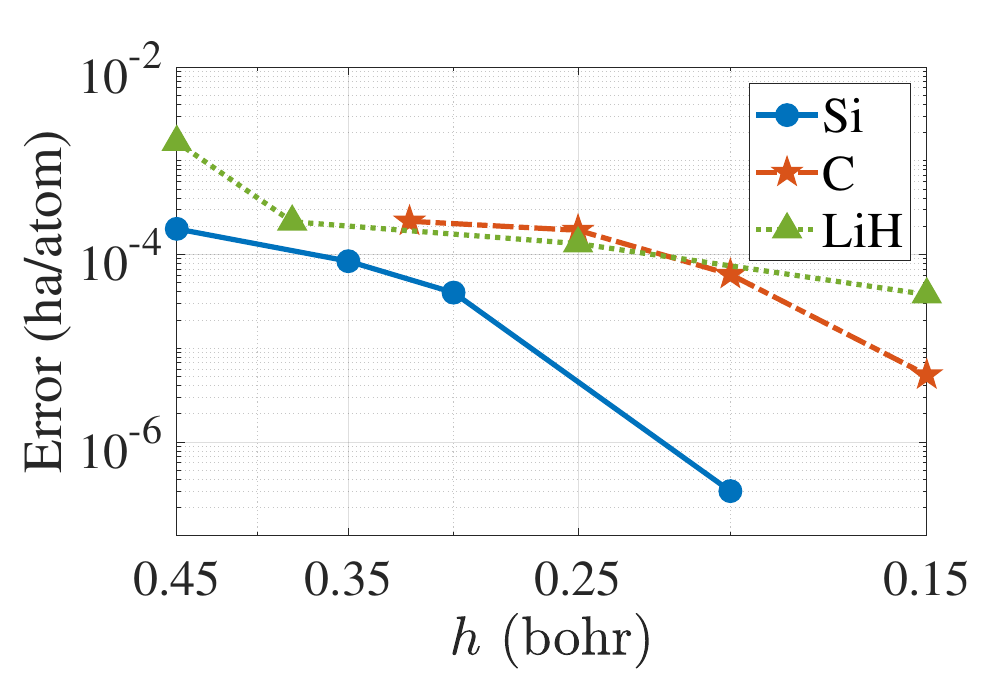} \label{fig:conv:mesh}} 
\caption{\label{Fig:Convergence} Variation of the RPA correlation energy with key parameters in the developed framework.}
\end{figure}

In Fig.~\ref{fig:conv:rank}, we plot the variation of the RPA correlation energy with respect to the normalized decomposition rank $N_r/N_e$. The error is defined with respect to the values obtained for $N_r/N_e = 350$. We observe rapid convergence in the RPA correlation energy, with  $N_r/N_e \sim 70$, $70$, and $30$ being sufficient to achieve chemical accuracy of $\sim 0.001$ ha/atom for the Si, C, and LiH systems, respectively. The corresponding numbers for an  accuracy of $10^{-4}$ ha/atom in the RPA correlation energy are $\sim 260$, $260$, and $170$, respectively. These results confirm that the low-rank assumption of the $\widetilde{\chi}_0$-matrix is a good approximation for the calculation of the RPA correlation energy. 

In Fig.~\ref{fig:conv:SternTol}, we plot the variation of the RPA correlation energy with respect to the Sternheimer tolerance $\varepsilon_s$. The error is defined with respect to the values obtained for $\varepsilon_s = 10^{-3}$. We observe that there is rapid convergence, with even a loose tolerance of $\varepsilon_s \sim 0.2$ sufficient to obtain chemical accuracy of $\sim 0.001$ ha/atom in the correlation energy. Notably, in recent work employing subspace diagonalization \cite{Shah2024Many}, the value of $N_r/N_e$ constrained how loosely the tolerance \( \varepsilon_s \) could be set, as the guess eigenvectors tended to become linearly dependent during the polynomial-filtered subspace iteration. However, this issue is not encountered in the current framework using SQ, which permits looser Sternheimer tolerances and therefore leads to significant speedups of up to a factor of two for the chosen systems.

In Fig.~\ref{fig:conv:mesh}, we plot the variation of the RPA correlation energy with the real-space grid spacing $h$. The error is defined with respect to the values corresponding to $h = 0.15$, $0.10$, and $0.10$ for the Si, C, and LiH systems, respectively. We observe that there is rapid convergence in the energy.  In particular, the correlation energy converges to within $10^{-4}$ ha/atom as the grid spacing is refined, achieving chemical accuracy of $\sim 0.001$ ha/atom with a grid spacing of $h \sim 0.45$ bohr. Notably, a similar grid spacing is required to achieve chemical accuracy for other choices of exchange-correlation functionals, confirming that an unnecessarily fine grid is not needed for the convergence of the RPA correlation energy.

In Table~\ref{table:Conv:SQOrder}, we present the variation of the RPA correlation energy with the SQ order $N_o$. We observe that there is rapid convergence in the energy. In particular, $N_o=4$ is sufficient to achieve an accuracy of $10^{-7}$ ha/atom for each of the systems. This can be attributed to the very small spectral width of the $\widetilde{\chi}_0$-matrix  when projected onto the subspace of the trial vectors, e.g., $\mathcal{O}(1)$ ha for the systems studied here, and the relatively smooth nature of the function $g(x) = \log(1-x) + x$ for which SQ is being performed \cite{suryanarayana2013spectral}.

\begin{table}
  \caption{Variation of the RPA correlation energy with SQ order $N_o$.}
  \centering
  \label{table:Conv:SQOrder}
  \begin{tabular}{cccc}
    \hline
     $N_o$ & 2 & 3 & 4\\
    \hline
    \ch{Si} & -0.2091340 & -0.2091346  & -0.2091347  \\
    \ch{C} & -0.2468244 & -0.2468245 & -0.2468246 \\
    \ch{LiH} & -0.0790038 & -0.0790044 & -0.0790045 \\
    \hline
  \end{tabular}
\end{table}


\subsection{Accuracy}
We next verify the accuracy of the developed framework through comparisons with the established planewave code ABINIT \cite{gonze2016recent}, which uses the direct approach for the calculation of the RPA correlation energy, i.e., construction of the $\nu \chi_0$ matrix, followed by an eigendecomposition. In particular, we consider a 2-atom cubic cell with dimensions of $5.14$, $3.36$, and $4.38$ bohr for the Si, C, and LiH systems, respectively, containing 8, 8, and 4 electrons. In SPARC, we employ a grid spacing of $0.20$, $0.15$, and $0.15$ bohr for the Si, C, and LiH systems, respectively, resulting in $N_d= \text{17,576}$, $\text{12,167}$, and $\text{27,000}$ grid points. In addition, we employ normalized decomposition ranks of $N_r/N_e = 350$, Sternheimer tolerance of $\varepsilon_s = 0.01$, SQ order of $N_o=4$, and subspace iteration correlation energy threshold of $\varepsilon_c = 10^{-5}$ ha/atom. With these choices, the  RPA correlation energies calculated by SPARC are converged to $\sim 10^{-4}$ ha/atom. In ABINIT, we use all unoccupied orbitals for the construction of the $\chi_0$-matrix.  In addition, we employ planewave cutoffs of $135$, $195$, and $165$ ha for the Si, C, and LiH systems, respectively. With these choices, the RPA  correlation energies calculated by ABINIT  are converged to $\sim 10^{-4}$ ha/atom. 

In Table~\ref{table:Comp:ABINIT}, we compare the RPA correlation energies computed by SPARC and ABINIT. We observe that there is very good agreement, with differences of $\sim 10^{-4}$ ha/atom, which are well within the desired chemical accuracy of $\sim 0.001$ ha/atom. Indeed, the agreement is expected to further increase on choosing larger planewave cutoffs in ABINIT, however such simulations  fail to execute on our computer cluster. Note that, based on the model developed in SPARC for the equivalent grid spacing in real-space calculations and the planewave cutoff in planewave calculations for the electronic ground state, the RPA correlation energy appears to converge faster in SPARC compared to ABINIT. This is likely due to the lower accuracy of the unoccupied states computed in ABINIT, as observed in the outputs. Also note that the current  implementation for the RPA correlation energy is applicable  to study molecules/clusters. However, we have not done so here because comparisons with planewave codes become even more challenging due to differences in boundary conditions and the large vacuum required, which significantly increases the degrees of freedom and prevents reaching the planewave cutoffs needed for a careful comparison.

\begin{table}
  \caption{RPA correlation energy computed by SPARC and ABINIT.}
  \centering
  \label{table:Comp:ABINIT}
  \begin{tabular}{cccc}
    \hline
     & ABINIT (ha/atom)& SPARC (ha/atom) &Difference (ha/atom)\\
    \hline
        \ch{Si}& $-0.20246$ &$-0.20235$& $0.9 \times 10^{-4}$\\
            \ch{C}& $-0.21899$ &$-0.21916$ & $1.7 \times 10^{-4} $\\
    \ch{LiH}& $-0.06300$ & $-0.06331$ & $3.1\times 10^{-4} $\\
    \hline
  \end{tabular}
\end{table}

\subsection{Performance}

We now study the performance of the developed framework for the calculation of the RPA correlation energy. In particular, we perform a strong scaling test, i.e., we study the variation time to solution as the number of processors is  increased, while holding the system size fixed. We consider $8$-, $32$-, and $64$-atom cuboidal cells of LiH with dimensions of $7.6 \times 7.6 \times 7.6$, $15.2 \times 15.2 \times 7.6$, and $15.2 \times 15.2 \times 15.2$ bohr, respectively, containing $16$, $64$, and $128$ electrons.  We employ a grid spacing of $0.40$ bohr, resulting in $N_d= \text{6,859}$, $\text{27,436}$, and $\text{54,872}$ grid points for \ch{(LiH)4}, \ch{(LiH)16}, and \ch{(LiH)32}, respectively. In addition, we employ a normalized decomposition rank of $N_r/N_e = 36$, Sternheimer tolerance of $\varepsilon_s = 0.1$, SQ order of $N_o=3$, and subspace iteration energy threshold of $\varepsilon_c =  10^{-4}$ ha/atom. With these choices, the  RPA correlation energy is converged to within chemical accuracy of $\sim 0.001$ ha/atom.  The simulations are performed on the \texttt{Phoenix} supercomputer at Georgia Institute of Technology, where each node has Dual Intel Xeon Gold 6226 CPUs @ 2.7 GHz (24 cores/node), DDR4-2933 MHz DRAM, and Infiniband 100HDR interconnect.

\begin{figure}
\centering
\includegraphics[keepaspectratio=true,width=0.49\textwidth]{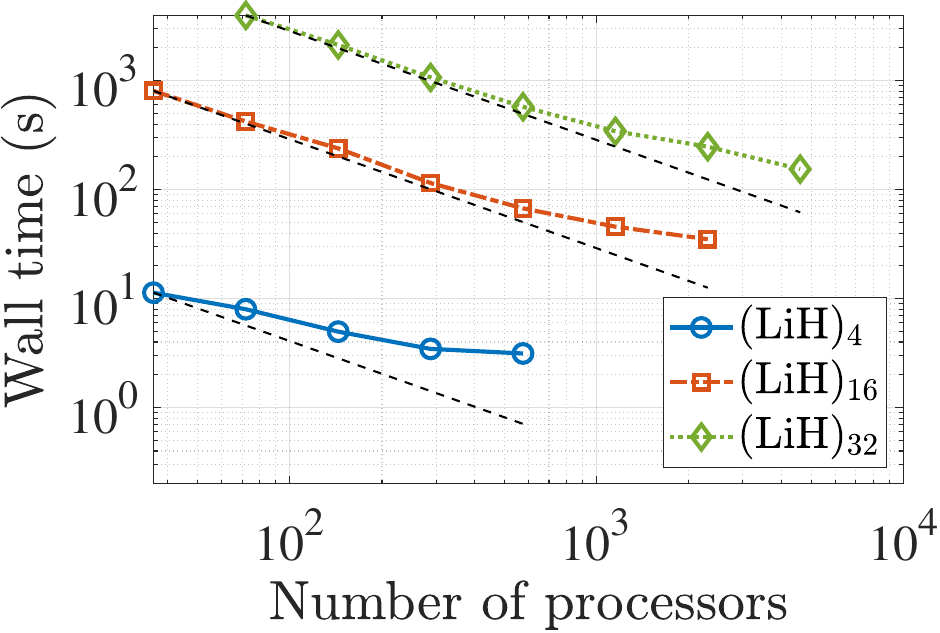} \label{fig:strongScaling}
\caption{\label{Fig:ParallelScaling} Strong scaling of SPARC for the calculation of the RPA correlation energy, where the dotted lines represent the ideal scaling.}
\end{figure}

\begin{figure}
\centering
\subfloat[\ch{(LiH)4}]{\includegraphics[keepaspectratio=true,width=0.44\textwidth]{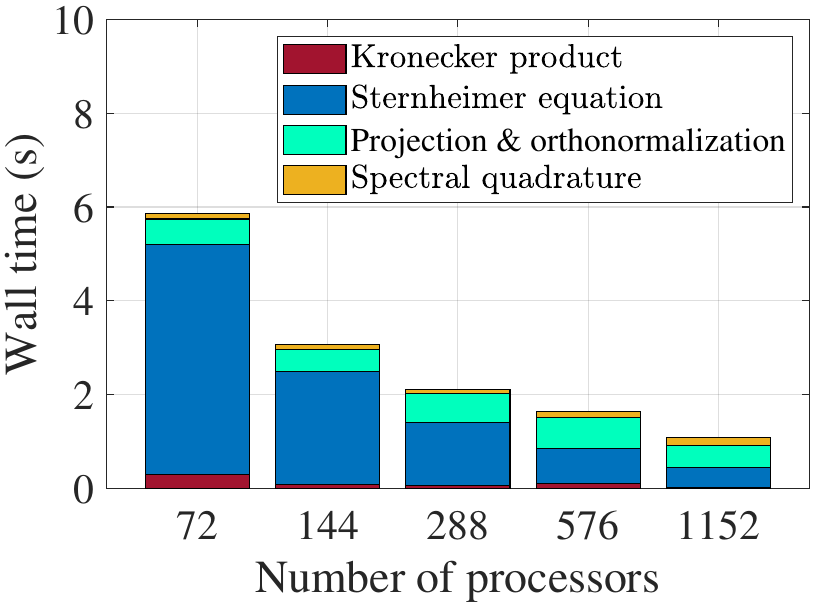} \label{fig:timingSplit16}} 
\subfloat[\ch{(LiH)16}]{\includegraphics[keepaspectratio=true,width=0.44\textwidth]{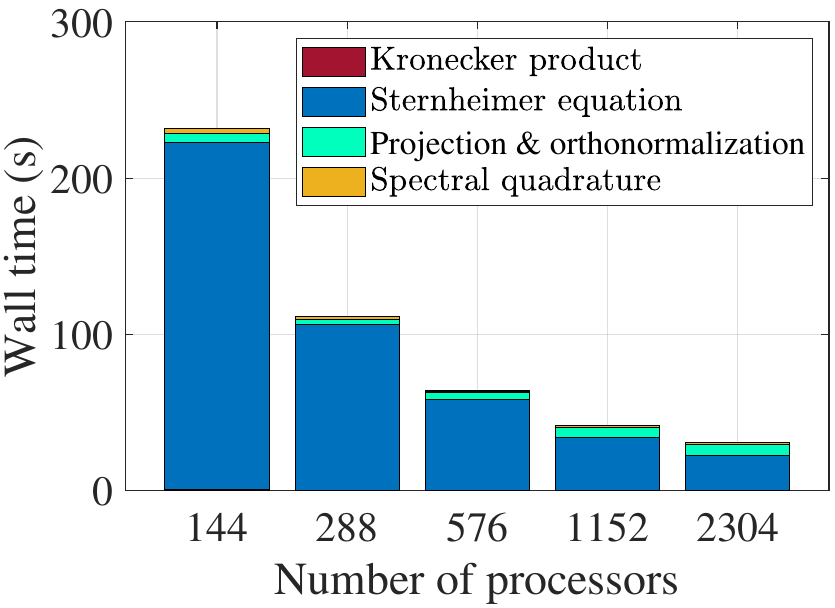}  \label{fig:timingSplit64}} \\
\subfloat[\ch{(LiH)32}]{\includegraphics[keepaspectratio=true,width=0.44\textwidth]{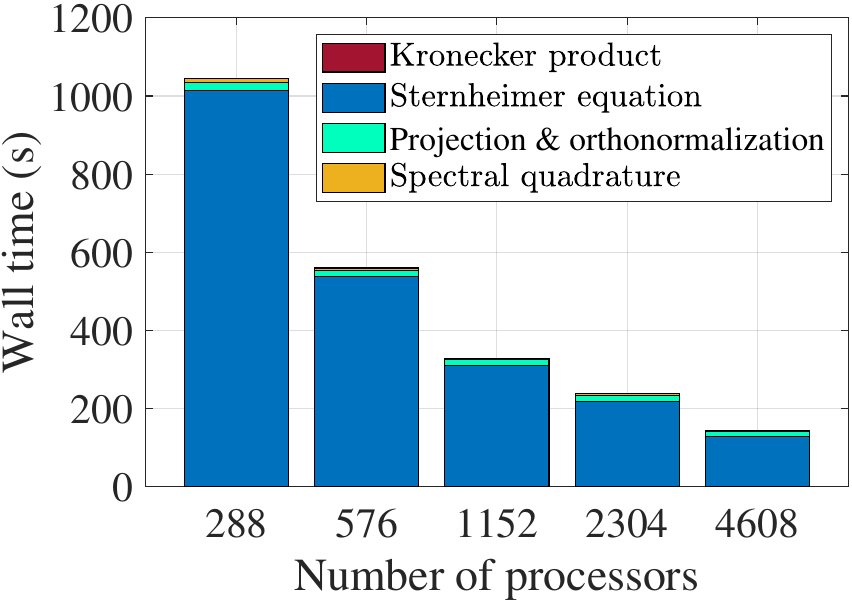}  \label{fig:timingSplit128}}  
\caption{\label{Fig:TimingSplit} Breakdown of the timings in the strong scaling study for the LiH systems.}
\end{figure}

We present the parallel strong scaling results so obtained in Fig.~\ref{Fig:ParallelScaling}. We observe that SPARC exhibits good scalability to thousands of processors, achieving efficiencies of 23\%, 36\%, and 40\% for the \ch{(LiH)4}, \ch{(LiH)16}, and \ch{(LiH)32} systems, respectively, on the largest processor counts of 576, 2304, and 4608. Indeed, higher efficiencies are expected due to the inherently parallel nature of solving the Sternheimer linear systems, which employs a two-level parallelization over both trial vectors and Kohn-Sham orbitals. However, as the number of processors is increased, the solution of the Sternheimer system becomes less dominant, and the scaling becomes limited by the other key steps, including projection of the $\widetilde{\chi}_0$-matrix onto the subspace spanned by the trial vectors and orthonormalization of the trial vectors. This is evident from Fig.~\ref{Fig:TimingSplit}, where the breakdown of the timings in the strong scaling study have been presented. In particular, on the smallest number of processors, the solution of the Sternheimer equations takes 83\%, 95\%, and 96\% of the total time, while on the largest number of processors, it occupies only 24\%, 63\%, and 82\% of the total time. In particular, the projection and orthogonormalization steps scale relatively poorly, which significantly reduces the overall scaling efficiency. Indeed, the time taken by the Kronecker product scheme and SQ are relatively minor fractions of the total time. Note that even within the time spent solving the Sternheimer equations, the generation of the initial guess --- computed simultaneously for all right-hand side vectors local to a processor --- exhibits poor strong scaling, as the \texttt{BLAS3} operations effectively reduce to \texttt{BLAS2} operations with increasing processor count. Indeed, excluding the time spent on generating the initial guess, the efficiency of solving the Sternheimer linear systems on the largest number of processors is 82.4\%, 86.5\%, and 86.4\% for the \ch{(LiH)4}, \ch{(LiH)16}, and \ch{(LiH)32} systems, respectively, with the loss in efficiency attributable to the load imbalance arising from the varying difficulty of the individual linear systems. We also observe from these results, based on the CPU times for the LiH systems at the smallest number of processors, i.e., 0.16, 9.56, and 85.70 CPU-hours for the \ch{(LiH)4}, \ch{(LiH)16}, and \ch{(LiH)32} systems, respectively, that the formalism scales nearly perfectly with the system size, i.e., $\mathcal{O}(N_d^3)$. 

In terms of comparison of the performance with the planewave code ABINIT, which uses the direct approach for the calculation of the RPA correlation energy, we consider the \ch{(LiH)4} system. We use a planewave cutoff of $75$ ha, which converges the RPA correlation energy to $\sim 0.001$ ha/atom, comparable to the accuracy of the SPARC results. The CPU time taken by ABINIT, which includes the electronic ground state calculation is $34$ hours, while the corresponding time taken by SPARC is $0.75$ hours, achieving more than an order-of-magnitude speedup. Given SPARC's cubic scaling formalism compared to ABINIT's quartic scaling, along with the greater parallel scalability of the DFPT-based approach, its advantages are expected to become more pronounced with increasing system size and the availability of more processors.


\section{Concluding Remarks}
In this work, we have presented a real-space method for computing the RPA correlation energy within Kohn–Sham DFT, leveraging the low-rank nature of the frequency-dependent density response operator. In particular, we have employed a cubic-scaling formalism based on DFPT that circumvents the explicit construction of the response function matrix, instead relying on the ability to calculate its product with a vector by solving the associated Sternheimer linear systems. We have developed a large-scale parallel implementation of this approach using the subspace iteration method in conjunction with the  SQ method, while employing the Kronecker product–based formalism for the application of  the Coulomb operator and the COCG method for the solution of the linear systems. We have demonstrated the convergence with respect to key parameters and verified the method’s accuracy by comparing with planewave results. In addition, we have demonstrated that the framework exhibits good strong scaling to many thousands of processors, reducing the time to solution for a lithium hydride system with 128 electrons  to around 150 seconds on 4068 processors.

The acceleration of the key computational kernels on GPUs, as implemented for local/semilocal \cite{sharma2023gpu} and hybrid  \cite{jing2025gpu} functionals is expected to significantly reduce the time to solution, making it a promising avenue for future research. Other worthwhile directions include extending the framework to incorporate Brillouin zone integration and exact exchange within the DFPT-based formalism, as well as developing a self-consistent formulation and large-scale parallel implementation for the RPA correlation energy.


\begin{acknowledgement}
The authors gratefully acknowledge the support of the U.S. Department of Energy, Office of Science under grant DE-SC0023445. This work was performed in part under the auspices of the U.S. DOE by LLNL under Contract DE-AC52-07NA27344. This research was also supported by the supercomputing infrastructure provided by Partnership for an Advanced Computing Environment (PACE) through its Hive (U.S. National Science Foundation through grant MRI1828187) and Phoenix clusters at Georgia Institute of Technology, Atlanta, Georgia.
\end{acknowledgement}


\providecommand{\latin}[1]{#1}
\makeatletter
\providecommand{\doi}
  {\begingroup\let\do\@makeother\dospecials
  \catcode`\{=1 \catcode`\}=2 \doi@aux}
\providecommand{\doi@aux}[1]{\endgroup\texttt{#1}}
\makeatother
\providecommand*\mcitethebibliography{\thebibliography}
\csname @ifundefined\endcsname{endmcitethebibliography}
  {\let\endmcitethebibliography\endthebibliography}{}

\end{document}